# Spreadsheet assurance by "control around" is a viable alternative to the traditional approach

*by Harmen Ettema, Paul Janssen and Jacques de Swart*

*PricewaterhouseCoopers*

*Amsterdam, The Netherlands*

*E-mail: harmen.ettema@nl.pwcglobal.com*

**ABSTRACT**

*The traditional approach to spreadsheet auditing generally consists of auditing every distinct formula within a spreadsheet. Although tools are developed to support auditors during this process, the approach is still very time consuming and therefore relatively expensive. As an alternative to the traditional 'control through' approach, this paper discusses a 'control around' approach. Within the proposed approach not all distinct formulas are audited separately, but the relationship between input data and output data of a spreadsheet is audited through comparison with a shadow model developed in a modelling language. Differences between the two models then imply possible errors in the spreadsheet. This paper describes relevant issues regarding the 'control around' approach and the circumstances in which this approach is preferred above a traditional spreadsheet audit approach.*

## 1 THE TRADITIONAL AUDIT APPROACH CAN BE VERY TIME CONSUMING

### 1.1 For moderately sized spreadsheets, traditional spreadsheet audit efforts are in the range of 25 to 75 hours

A moderately sized spreadsheet typically involves thousands of cells. Spreadsheet Audit (SA) tools like OAK [1] and Spreadsheet Professional [2], reduce the audit to the inspection of the distinct formulas only. The number of distinct formulas strongly depends on the nature of the spreadsheet, but for moderately sized spreadsheets this number typically lies in the range of 500 to 1,500. Besides the mentioned SA tools, features in spreadsheet programs like "trace dependents" and "trace precedents" in Microsoft Excel, help the spreadsheet auditor to validate distinct formulas quickly. But still, as a rule of thumb, the inspection of a distinct formula by an experienced auditor takes on average around 3 minutes. The total effort of such a traditional spreadsheet audit, or in assurance terms control through approach, generally takes from 25 up to 75 hours. This amount of time is often out of proportion compared to the development of the spreadsheet. On the other hand, in many cases the assurance of the spreadsheet is still required by users or third parties, which decide to invest large sums of money based on the spreadsheet calculation.

### 1.2 For large spreadsheets, traditional SA often is not an option because throughput time is too long

If the number of distinct formulas is much larger 1,500, then not only the time spend and the audit costs are an obstacle for following the traditional SA approach, but also the throughput time. For complicated, badly structured spreadsheets, it can be hard to perform the SA by more than one auditor in parallel. An example of such a complicated, badly structured spreadsheet is a spreadsheet with on one worksheet thousands of distinct and repetitive formulas. Consequently, the throughput time for these spreadsheets will be, based on the previous assumptions, more than two weeks. Because users want to take the spreadsheet in production as quick as possible the time-pressure on the SA is very high, and therefore often limited to one or two weeks.

## 2 THE CONTROL AROUND APPROACH EXPLOITS ADVANTAGES OF MODELLING LANGUAGES

### 2.1 The control around approach relies on a shadow model that validates output in the spreadsheet

In our proposed audit approach, we build a shadow model of (parts of) the spreadsheet model in a modelling language. The shadow model consists of the formulas in the spreadsheet that need to be audited. By importing scenarios of input data in the spreadsheet as well as in the shadow model, one can compute various scenarios according to the shadow model and compare it to the output in the spreadsheet. These scenarios should reflect a relevant set of cases of input data for the spreadsheet model. Defining the right test set is crucial in a shadow modelling approach. By focusing on the input/output relationship of data the validation of all intermediate calculations in the spreadsheet can be restricted to differences found between output of the spreadsheet and the shadow model. In appendix A we describe the basic features of this modelling language, which is integrated in a modelling environment called AIMMS [6]. Appendix B illustrates the building of such a model using a simple sales forecasting example in the field of telecommunication. In the remainder of this section, we highlight the advantages of using modelling languages for SA.

### 2.2 Defining the right test set of input data scenarios is crucial for a successful control around approach

Assurance on the spreadsheet is given by comparing the output of the spreadsheet with the output of the shadow model for a test set of input data scenarios. Therefore it's crucial to compose a relevant test set. The test set can for example consist of a scenario with default values, scenarios with mutations in values for individual input variables and scenarios with mutations in values for combinations of input variables. A test set can also consist of scenarios with random numbers between possible values for each input variable. A way to overcome the disadvantage of a limited test set is to perform parallel processing for important calculations. This means that the output is not only produced with the spreadsheet model but also with the shadow model.

### 2.3 Separation of data and formulas is a key advantage in a modelling language

Although a decently designed spreadsheet should make a clear distinction between the regions in the spreadsheet that contain data manipulation and those that contain pure data, many spreadsheets encountered in practice lack this property. The separation of data and relationships is the cornerstone of a modelling language, in which only symbolical relations are defined, and the loading of the input data into the model parameters is a separate process (see Appendix A). Another property of properly built spreadsheets is that they use named ranges that make the formulas more readable and easier auditable. In practice, one often violates this design rule, because it is too time consuming to update the range names after adjusting the spreadsheets. In a modelling language, the formulas typically take the form of verbose English phrases that are easy to audit, because they are even understandable for people who are not familiar with the concept of modelling languages. An adjustment of the input data typically does not require the adjustment of the formulas.

### 2.4 Top-down approach is a logical consequence of a modelling language

Most spreadsheet builders try to follow some sort of top-down approach. This often comes down to arranging the workbook so that each worksheet contains a logical functional module and that data is "pushed" to the right and down through cells and worksheets within the spreadsheet. It is clear that such an approach greatly simplifies the work of the auditor of the spreadsheet. However, adjustments to the spreadsheet often lead to violating this principle. In a modelling language the model is of a hierarchical nature, which results in clear insight into the dependency structure.

### 2.5 Incremental software development approaches are hard to establish in a spreadsheet, but easy in a modelling language

#### 2.5.1 Rapid Application Development is the state of the art incremental approach

Rapid Application Development (RAD) has become a popular way of designing computer applications in the past decade. Before RAD became the trend, one typically first made extensive, time-consuming specifications of what to build. The programming was only started after the approval of the specification. However, the shortcomings of the specification often appeared after the end-user had worked with the first version of the software. To deal with these shortcomings, one started the iteration loop *specification adjustment – specification approval – programming – inspection by end-user*. This iteration turned out to be very time consuming. RAD focuses on a rapid development of a first version that only contains the basic features of the application. The end-user inspects this version, comments on the basic features and indicates with which other features the application should be enriched. Then the iteration *adjustment of existing features and programming of new features – inspection by end-user* is started. In this iteration, the specification of the application, which is often difficult to understand for the end-user, is only used for internal use by the programmer. For a more elaborate discussion of the advantages of RAD, we refer to [3].

For SA, the use of RAD can be very beneficial, because in the first iteration one can focus on the building of a shadow model that computes the key output data in the spreadsheet. This could be done for a small data set only. For example, if the spreadsheet involves a financial forecast for several countries for several years, one could start to establish a shadow model for the first year and one country only. In the second iteration the data set could be enlarged and the computation of other, less important output data could be added to the shadow model.

#### 2.5.2 Rapid Application Development is hard to establish in a spreadsheet

Due to the simple user interface, spreadsheets are easy to make. Even without or with only limited training it is possible to develop large spreadsheets with many features. However, adjustments and supplements, which are the basic ingredients of the RAD approach, are relatively hard/risky to incorporate in spreadsheets. An adjustment often requires the error-prone and time-consuming revision of many other cells. Adding an element to a dimension, e.g. adding a additional year to the time horizon, typically involves inserting additional rows or columns. As a consequence, all time dependent formulas and graphs have to be extended.

#### 2.5.3 Rapid Application Development is a natural concept in a modelling language

Due to its hierarchical structure, a modelling language is well suited for RAD, because it is easy to add branches to the model structure without adapting other branches. Enlarging a dimension is as easy as adding new records to database tables. Adding a dimension is equivalent to adding a index to all relevant parameters. E.g., making the model time dependent is just a matter of adding a time index to all relevant parameters, whereas for a spreadsheet, this might lead to rebuilding the spreadsheet from scratch. Since the Graphical User Interface (GUI) is orthogonal to the model, new features in the GUI do not interact with the model structure and are therefore not subject the model assurance. For more details on RAD in a modelling language, we refer to Appendix A.

### 2.6 Control around approach exploits a library of common formulas

#### 2.6.1 Even common relationships have to be audited again for every spreadsheet

A large fraction of a typical spreadsheet consists of a set of common relationships which are used in other spreadsheets as well. For example, many spreadsheets use the same financial forecast formulas. However, since the data structure is different per spreadsheet, the auditor has to check these common formulas again for every new spreadsheet. This makes the SA work along the traditional approach not only quite boring, but also inefficient.

### 2.6.2 The shadow model is built using a library of well-tested, robust, state of the art, common relationships

Instead of checking the same formulas over and over again, an auditor who uses the control around approach typically builds these common formulas for the shadow model once. In a modelling language these formulas are in readable English and are therefore easily confirmed using a standard text book from the literature. The auditor collects these formulas in a library. This library can be linked to the shadow model for the new spreadsheet that has to be audited. At PricewaterhouseCoopers, we already have a wide collection of frequently occurring formulas, which makes the formulation of shadow models very efficient. In many cases the library formulas can be used directly. In the worst case, the data structures in the spreadsheets are so different, that the auditor has to change the index domain of the parameters occurring in the formulas. However, this does not affect the basic structure of the formula.

## 3 ESPECIALLY FOR HIGH-DIMENSIONAL MODELS THE CONTROL AROUND APPROACH IS BENEFICIAL

### 3.1 High dimensional problems create many distinct formulas in a spreadsheet

A spreadsheet is most convenient for two-dimensional models. The first dimension uses the rows, the second the columns. If the model contains a third dimension, then the first option is to create worksheets that contain individual slices of the third dimension. Relationships that depend on three dimensions are from a modelling perspective not more complex than those that depend on two dimensions. However, in a spreadsheet every slice of the third dimension generates a distinct formula, which has to be audited separately. Therefore, the larger the cardinality of the third dimension, the less efficient the traditional SA approach.

A formula that aggregates over the third dimension needs to contain references to all individual worksheets. For example, if the third dimension represents time in years, then every worksheet represents one year. The formula for the Net Present Value (NPV) would be a long expression with references to many sheets.

Another option for storing the third dimension is generating separate blocks on the same sheet, each block representing a slice of the third dimension. This option has to be selected if the model involves more than three dimensions. Again formulas with parameters defined over one or more dimensions that are represented by different blocks, generate more than one distinct formula.

### 3.2 In a modelling language every relationship is represented by one distinct formula independent of dimensions

Instead of auditing all distinct formulas in a spreadsheet used to represent one single relationship, the control around approach proposes to replace all these formulas by one formula in the shadow model. After feeding the input data from the spreadsheet in the shadow model, the shadow model is evaluated and its output compared with the output in the cells of the spreadsheet. This comparison process is often efficient, because it normally only involves the key output data, which are at an aggregated level.

Based on the previous discussion, it is possible to compare the efficiency of the traditional approach with the control around approach. Suppose that the model has $d$ dimensions, and that every dimension has the same number of levels. Furthermore, we assume that, due to the copying of the data to the modelling environment, the cost of auditing one relationship via the control around approach is twice as expensive as auditing one distinct formula via the traditional approach. The following graph shows the average effort of both approaches for auditing one relationship that involves all dimensions versus the number of levels for $d =$ 1,2,3 and 4, under ceteris paribus conditions. From this graph it is clear that for higher values of $d$, the control around approach becomes less time consuming than the traditional approach for increasing values of the number of levels per dimension.

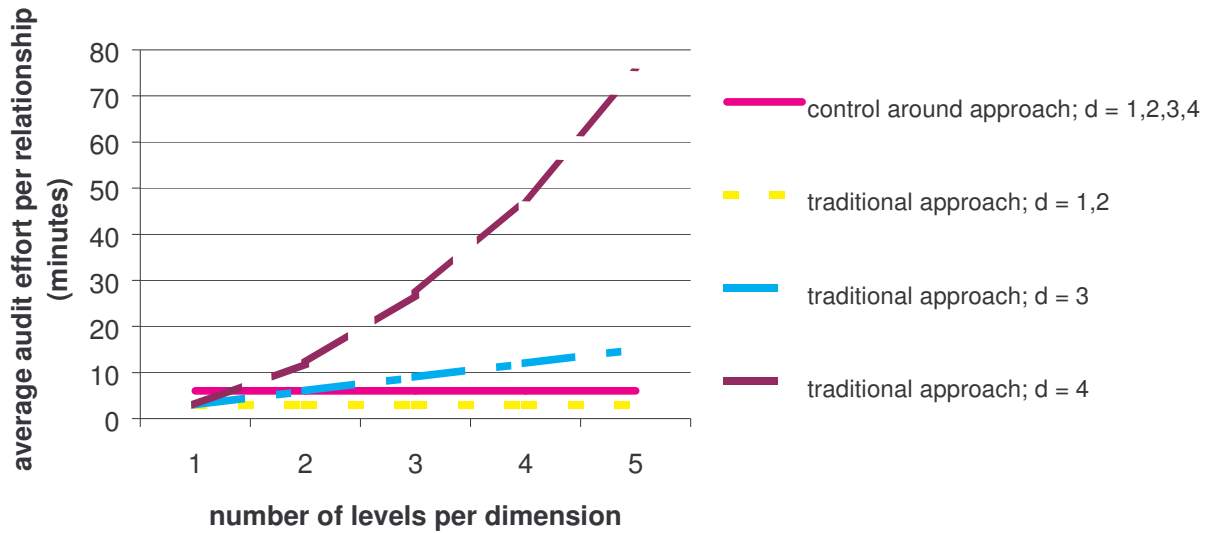

## 4 THE CONTROL AROUND APPROACH DELIVERS MORE THAN JUST ASSURANCE

### 4.1 The shadow model is a good starting point for strategic scenario analysis

Once the spreadsheet has been audited, one may decide to use the shadow model for the further development of the model. An interesting enhancement of a model is the study of the impact of different business strategies on the key performance indicators under several scenarios. In the field of Operations Research one often formulates an optimisation problem, in which the task is to find the optimal setting of the decision variables (the optimal business strategy) for different settings of the environmental variables. Once the model has been formulated in a modelling language, such optimisation problems can be easily defined and solved using state of the art solver like CPLEX [4] and CONOPT [5], which are available in the modelling environment.

### 4.2 The shadow model can be visualised efficiently

The Graphical User Interface of the modelling environment is well-suited for a quick development of a natural user interface. Moreover, it is easy to create a pdf report containing all conclusions and intermediate output.

### 4.3 The control around approach makes SA more exciting than the traditional approach

Instead of the tedious, monotone and often boring work involved in the traditional SA approach, the control around approach offers a more challenging type of work, for which it is required to think about the structure of the model on a conceptual level.

### 4.4 The control around approach gives an overall assurance, while traditional assurance is focused on accuracy of formulas

The functionality of the spreadsheet is rebuilt in a shadow model, based on the spreadsheet, conversations with spreadsheet users and available documentation. With this shadow model the spreadsheet is audited.

Therefore, the control around approach gives an overall assurance on the relevant input/output relationships as covered by the test set, while the traditional method is focused on individual calculations. The extra element of interpreting the formulas, rebuilding the model and comparing the results gives not only assurance on correct formulas but also assurance on the accuracy of the model as a whole: does the model do what it is supposed to do.

### 4.5     Should spreadsheets be used at all?

After presenting the list of advantages of using the control around approach above the traditional SA approach for complex multidimensional spreadsheets it's justified to ask the question if we should use spreadsheets at all? It is clear that from the perspective of modelling strength we prefer modelling languages above spreadsheets. The separation of data and formulas, the hierarchical structure of the formulas and the scalability and therefore ability for incremental software development methods are key advantages of modelling languages. Although the strength of modelling is of course just one of the software selection criteria, the importance of this criteria is often underestimated. For complex models the modelling strength of the underlying software is the key success factor for the prevention of modelling and user errors. Software for supporting company critical decisions should therefore be reviewed twice on their modelling strength.

## APPENDIX A: BASICS OF MODELLING LANGUAGES

In this appendix we describe a few basic features of the modelling language AIMMS [6]. Other modelling languages have similar features. For a more elaborate description, we refer to [6].

In translating real-life problems into AIMMS models, several conceptual steps are required. First, the input and output data are described using *sets* and *indexed parameters*. *Sets* are entities that determine the size of the problem. For example, in a transport problem, the cities between which transport takes place form such an entity. With these entities a number of instances is associated. For example, one of the cities could be 'Amsterdam'. However, to keep the model generic and maintainable, it is customary and desirable to translate the problem into a symbolic model and not to make any explicit reference to individual instances. Such high-level model specification can be accomplished through the use of sets, each with an associated index for referencing arbitrary elements in that set. AIMMS allows for hierarchical set structures. E.g., origin cities and destination can be subsets of one set `Cities`. *Indexed parameters* are used to store data that can be associated with a particular element or tuple of elements. For example, if `o` is an index in the set `Origins` and `d` is an index in the set `Destinations`, then the distances between origin cities and destination cities can be stored in the parameter `Distance(o,d)`. Data can be fed into the input parameters via the AIMMS Graphical User Interface (GUI) or via links with ASCII files, databases, spreadsheets or internal procedures that construct the input data.

Second, the mapping between input and output parameters is described using AIMMS definitions or procedures. Procedures are used for output parameters that cannot be expressed in terms of other parameters in closed form. For example, there is no closed form for the Internal Rate of Return (IRR). AIMMS definitions can be entered together with the declaration of the output parameters in a concise and verbose form and are often powerful enough to describe very complex relationships. As an example we give the definition of the largest national distance: `LargestNationalDistance := MAX((o,d) | Country(o) = Country(d), Distance(o,d))`. Here, the input parameter `Country(c)` specifies in which country the city is situated, where `c` is an index in `Cities`, and the expression `(o,d) | Country(o) = Country(d)` assures the maximum is taken over distances between cities which lie in the same country.

Third, the model is executed. For output parameters that are computed via a procedure, this means that a call to this procedure has to be made. Parameters with a definition are automatically evaluated as soon at least one input parameter on which the definition (in)directly relies is changed.

The AIMMS model can be defined in a point-and-click manner via a GUI, which assures that the user has to learn hardly any syntax.

## APPENDIX B: EXAMPLE OF SA BY CONTROL AROUND

The following case is based on a client case. For the purpose of illustration and competition considerations, the data is fictitious and some case elements have been strongly simplified.

Consider a company that studies the possibility of investing in a data telecommunication network. To assess the feasibility of this project, the Future Cash Flows (FCFs) generated by the network have been modelled as follows: For the next 20 years, estimates of the volume of communications over the network for three different scenarios (worst case, base case and best case) have been made. In the proposed business model, users of the network pay a fixed amount per transmitted byte, independent of the trajectory in the network. However, due to regional differences in tax legislation, the gross margin for the company depends on the specific trajectory. The major investments in the network are the acquisition and engineering costs for the glass fibre cables. These investments are done in the first year and come down to a fixed price per mile. For simplicity of illustration, we neglect other investments here.

The company has constructed a spreadsheet to implement this model. In this spreadsheet, the communication network is depicted as a matrix in which the rows represent the origins and the columns the destinations. One worksheet contains a distance matrix[1], a matrix that specifies the gross margin as function of the trajectory in the network and the following constants: investment costs per mile, Weighted Average Costs of Capital (WACC) and the sales price per byte of transmitted data. For every year in the time horizon of 20 years, a separate worksheet contains the three scenarios for the communication volumes. In the same worksheet, a formula is included that computes the net revenues per scenario for the corresponding year. The Net Present Value (NPV) is computed from a formula, which involves cells from all of these 21 worksheets, in a separate results worksheet.

Since one of the most important Key Performance Indicators to the company is the Net Present Value (NPV), the SA focuses on auditing the formulas for the NPV. Using the traditional SA approach, the auditor first has to validate the formulas for the net revenues for each year individually. Next, he/she has to validate whether the NPV formula contains the correct references to the individual worksheets. Using our proposed control around approach, we build a shadow model as follows. First, we identify four dimensions: time (t), Origin (o), Destination (d) and Scenario (s). Second, we declare the following parameters: `Volume(o,d,t,s), GrossMargin(o,d), Distance(o,d), InvestmentPerMile, WACC, PricePerByte` and link the data in the corresponding matrices and cells in the spreadsheet to these parameters. Third, we define the following parameters:

- `Investment(t) := IF (t = FIRST(Time)) THEN SUM((o,d), Distance(o,d) * InvestmentPerMile ELSE 0 ENDIF;`
- `FCF(s,t) := Volume(o,d,t,s) * GrossMargin(o,d) – Investment(t);`
- `NPV(s) := SUM(t, FCF(s,t)/(1 + WACC)^(t – 2001)));`

Finally, we compare the data in NPV(s) with the three corresponding cells in the results worksheet.

From this example we see that in the control around approach, there are only three formulas, which are easy to audit due to their verbose nature, as opposed to at least 23 distinct formulas in the traditional approach, which are harder to audit because of their less verbose nature and considerable size (e.g. for the NPV). Furthermore, the visualisation features in AIMMS make it easy to draw the network in the form of a graph, in which the origins and destinations are represented by nodes, and the trajectories by arcs. Especially when the geographical coordinates of the origins and destinations are known and used for positioning the nodes of the graph, this type of visualisation is very powerful. An example of such a Graphical User Interface (GUI) is shown in the figure below. In order to inspect the model data further, the user can select for which year and which scenario the data is displayed along the arcs of this graph via the drop down lists in the upper right corner.

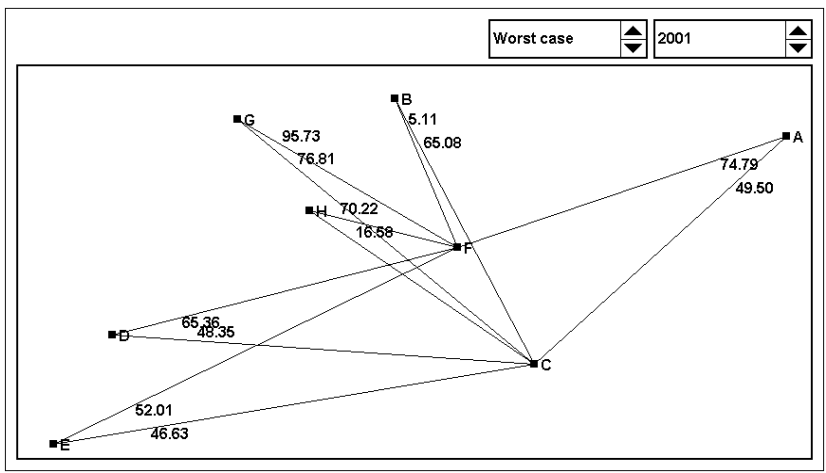

---

[1] In the distance matrix, only cells that refer to direct trajectories, i.e. trajectories which do not involve other points in the network than the origin and the destination, are non-empty. Furthermore, to avoid double counting, the matrix is lower triangular.